\documentclass[prb]{revtex4}
\usepackage{graphicx}
\begin{document}

\title{Orbital ordering and frustrations}
\author{D. I. Khomskii and M. V. Mostovoy}
\affiliation{Materials Science Center, University of Groningen, Nijenborgh 4, 9747 AG Groningen, The Netherlands}

\begin{abstract}

An orbital ordering occurs in many transition metal compounds with Jahn-Teller ions (Cu$^{2+}$, Mn$^{3+}$, low-spin Ni$^{3+}$,
Ti$^{3+}$ etc.). It plays an important role in these materials. At the same time, exchange interactions in orbitally degenerate systems are inherently frustrated, even in materials with simple crystal lattices. We discuss the origin of this frustration, considering in detail materials with a cubic and triangular  lattice of transition metal ions. We also discuss possible types of ground states of such systems, e.g., disordered orbital liquids and ordering due to the order-from-disorder mechanism.

\end{abstract}

\maketitle

\section{Introduction}

Many transition metal (TM) compounds show an orbital ordering \cite{KugelKhomskii,TokuraNagaosa,vdBrinkKhaliullinKhomskii}. Typical examples are TM oxides with so-called strong Jahn-Teller (JT) ions, e.g. Cu$^{2+}$(d$^9$), Mn$^{3+}$(d$^{4}$), low-spin Ni$^{3+}$(d$^{7}$) etc, which in the cubic crystal field of a regular oxygen octahedron have one electron or one hole on the doubly degenerate $e_g$ level. According to the well-known
Jahn-Teller theorem, such states are inherently unstable, and at low temperatures a distortion of the local coordination lifts this orbital degeneracy. In concentrated systems it leads to a phase transition - the so-called cooperative Jahn-Teller effect or orbital ordering \cite{KugelKhomskii}. An orbital ordering is also observed in materials with partially
filled three-fold degenerate $t_{2g}$-levels of TM ions, e.g. Ti$^{3+}$($d^1$) or V$^{3+}$($d^2$). Some of these materials have very unusual properties \cite{Ren,Keimer,KhaliullinMaekawa}. However, the JT effect for  $t_{2g}$-electrons is usually weaker than for $e_g$-electrons and can be strongly affected by the spin-orbital interaction $\lambda {\bf L} \cdot {\bf S}$, which  for $e_g$-electrons is essentially quenched. This complicates the discussion of orbital ordering in $t_{2g}$-systems and in this paper we concentrate on materials with $e_g$ electrons, though many of our conclusions can be applied to $t_{2g}$-systems too.

\begin{figure}
\centering
\includegraphics[width=4cm]{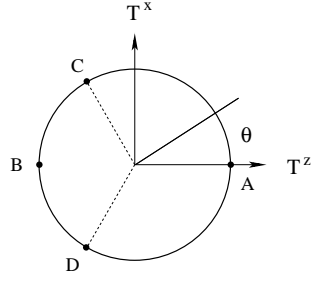}
\caption{\label{circle} The $(T^z,T^x)$ pseuodospin plane. The points A, B, C and D on the circle correspond, respectively, to the $d_{3z^2-r^2}$, $d_{x^2-y^2}$, $d_{3x^2-r^2}$, and $d_{3y^2-r^2}$ orbitals.}
\end{figure}

An orbital ordering can be described similarly to a magnetic (spin) ordering. In the case of one electron or hole on a doubly-degenerate orbital, one can describe the orbital occupation introducing a pseudospin (or isospin) $T = \frac12$, so that, e.g., the $d_{3z^2 -r^2}/d_{x^2-y^2}$ orbital
corresponds to the state with $T^z = +\frac12/T^z = - \frac12$.
A superposition of these two orbital states
\begin{equation}
|\theta\rangle = \cos \frac{\theta}{2} |3z^2-r^2 \rangle + \sin
\frac{\theta}{2} | x^2 - y^2\rangle, \label{eq:super}
\end{equation}
is associated with a particular point on a circle in the  $(T^z,T^x)$-plane, see Fig.\ref{circle}. We only consider here linear combinations with real coefficients. In principle, one can also consider states with complex coefficients, e.g.,
\[
|\pm\rangle = \frac{1}{\sqrt{2}}
\left(|3z^2-r^2\rangle \pm i |x^2 - y^2\rangle\right),
\]
but they are rather special: they do not couple to the lattice, but break the time-reversal symmetry and have a nonzero magnetic octupole moment \cite{Khomskii}. Such states are eigenstates of the ${\bf T}^y$ operator that usually does not appear in the exchange Hamiltonian of doubly-degenerate JT systems.

The pseudospin operators can be used to describe interactions between orbitals on different TM sites in the same way one describes the spin exchange. An orbital ordering corresponds
then to a nonvanishing average pseudospin $\langle T_i \rangle$.  However, the orbital exchange usually has a more complicated form than the spin exchange, due to the directional nature of orbitals. The strong anisotropy gives rise  to a frustration of orbital ordering even in materials with simple bipartite lattices.

\section{Phonon-mediated orbital exchange}

We illustrate this point by considering effective orbital interactions in a perovskite lattice -  a simple cubic lattice of TM ions with oxygens located on the edges between them (we ignore here for simplicity the tilting of (TM)O$_6$-octahedra present
in many materials, which lowers the symmetry from cubic to orthorombic). There are two mechanisms of orbital ordering: a  lattice instability due to the JT coupling of electrons to lattice \cite{KugelKhomskii,GehringGehring,EnglmanHalperin}, and a purely electronic mechanism resulting
from the virtual hopping of TM electrons between neighboring sites \cite{KugelKhomskii2,KugelKhomskii}. Both lead to essentially the same type of effective exchange interactions. We first discuss the lattice-mediated orbital exchange.

Consider a pair of neighboring TM sites $j$ and $j+z$ along the $z$ axis (see Fig.\ref{jt}). For this pair the electron-lattice interaction responsible for lifting the double degeneracy of the $e_g$-states has the form
\begin{equation}
H_{JT} = - g \sum_{i=j,j+z} \left(d_{i1\sigma}^\dagger
d_{i1\sigma} - d_{i2\sigma}^\dagger d_{i2\sigma} \right)
Q_{3i} + \frac{K Q_{3i}^2}{2}, \label{eq:JT}
\end{equation}
where $d_{i1\sigma}^\dagger$ creates the
state $|3z^2-r^2\rangle$ at the site $i$ with the spin projection $\sigma$ and $d_{i2\sigma}^\dagger$ is the creation operator of the other $e_g$-state $|x^2 -y^2\rangle$. The $Q_{3i}$ phonon coordinate is the linear combination of the shifts of oxygens (from the O$_6$-octahedron containing the TM site $j$) along the corresponding TM-TM bonds (see Fig.\ref{jt}):
\[
Q_{3i} = \frac{1}{\sqrt{6}} \left[ 2 \left(u_{j+z/2} - u_{j-z/2}\right) - \left(u_{j+x/2} - u_{j-x/2}\right) - \left(u_{j+y/2} - u_{j-y/2}\right) \right].
\]

\begin{small}

\begin{figure}
\centering
\includegraphics[width=3cm]{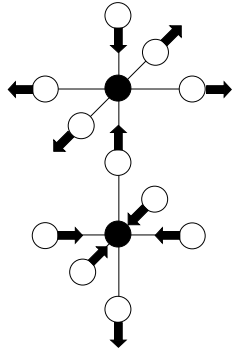}
\caption{\label{jt} A pair of TM ions (black circles) along the $z$ axis in the perovskite lattice.  Arrows indicate the shifts of oxygen ions (white circles) resulting in lifting of the double degeneracy present for a regular O$_6$-octahedron.}
\end{figure}

\end{small}

In the subspace of states with one electron per TM site the
combination of the electron operators $d_{i\alpha\sigma}^\dagger$ and $d_{i\alpha\sigma}$ in Eq.(\ref{eq:JT}) is equivalent to the pseudospin operator:
\begin{equation}
\sum_{\sigma} \left(d_{i1\sigma}^\dagger d_{i1\sigma} - d_{i2\sigma}^\dagger
d_{i2\sigma} \right)\equiv  2T^z_i.
\end{equation}
Excluding the coordinate $u_{j+z/2}$ of the oxygen located between the two TM ions, we obtain for an effective
orbital interaction for the pair
\begin{equation}
H_{j,j+z} = J T^z_j T^z_{j+z},\;\;\;\;\; J = \frac{2g^2}{3K},
\label{Hz}
\end{equation}
which has the (antiferroorbital) Ising form favoring the antiparallel alignment of neighboring pseudospins, e.g.
$T^j_1= +\frac12$ (the $d_{3z^2-r^2}$ orbital on the site $j$) and $T^z_{j+z} = - \frac12$ (the $d_{x^2 - y^2}$ orbital on site $j+z$).

However, complications arise when we consider effective orbital interactions in the whole cubic lattice of TM ions. Since in the cubic  lattice the three directions $x$, $y$, and $z$ are equivalent, we can describe the exchange along the $x$ direction by replacing the basis of states ($|3z^2 -
r^2 \rangle$, $|x^2-y^2\rangle$) by ($|3x^2 - r^2\rangle$, $|y^2-z^2\rangle$) and similarly for the $y$ direction. The symmetry of the cubic lattice under circular permutations of the $x$, $y$, and $z$ axes is reflected in the three-fold symmetry of the $(T^z,T^x)$-plane (see Fig.\ref{circle}), where the states $d_{3z^2-r^2}$, $d_{3x^2-r^2}$, and $d_{3y^2-r^2}$ correspond, respectively, to $\theta = 0$ (point A in Fig.\ref{circle}) , $\theta = +\frac{2\pi}{3}$ (point C), and $\theta = - \frac{2\pi}{3}$  (point D), where
\[
|\theta = \pm
\frac{2\pi}{3}\rangle = - \frac{1}{2} |3z^2 - r^2\rangle \pm
\frac{\sqrt{3}}{2} |x^2 - y^2\rangle.
\]
One can also introduce the corresponding symmetric notation for pseudospin operators
\begin{equation}
I^z = T^z, \;\;\;\; I^{x,y} = - \frac{1}{2} T^z \mp
\frac{\sqrt{3}}{2} T^x,
\label{Ioperators}
\end{equation}
so that $I^{\gamma} |3\gamma^2 - r^2 \rangle = + \frac{1}{2} |3\gamma^2 - r^2 \rangle$, where $\gamma = x,y,z$. Using the symmetry of the cubic lattice and Eq.(\ref{Hz}), we can then   immediately write the Hamiltonian of the exchange along the $x$ axis,
\begin{equation}
H_{j,j+x} = J I_j^x I_{j+x}^x = J
\left(\frac{1}{2}T^z_i+\frac{\sqrt{3}}{2}T^x_i\right)\left(\frac{1}{2} T^z_{j} + \frac{\sqrt{3}}{2} T^x_j\right),
\label{Hx}
\end{equation}
and similarly for a pair along the $y$ axis:
\begin{equation}
H_{j,j+y} = J I_j^y I_{j+y}^y = J
\left(-\frac{1}{2}T^z_i+\frac{\sqrt{3}}{2}T^x_i\right)\left(-\frac{1}{2}
T^z_{j} + \frac{\sqrt{3}}{2} T^x_j\right).
\label{Hy}
\end{equation}

The full Hamiltonian of lattice mediated effective orbital interactions has the form
\begin{equation}
H = J \sum_{i} \left(I_i^x I_{i+x}^x + I_i^y I_{i+y}^y +
I_i^z I_{i+z}^z \right) - \frac{3}{2} J \sum_i \left( \left(T^z_i\right)^2 + \left(T^x_i\right)^2\right).
\label{isocompass}
\end{equation}
It includes also the on-site term (the second sum in Eq.(\ref{isocompass})), which suppresses complex orbitals discussed above by favoring alignments of pseudospins in the $(T^x,T^z)$ plane.

We can now see the source of the frustration in the orbital sector: while the exchange along the $z$ direction stabilizes alternating $d_{3z^2 - r^2}$ and  $d_{x^2-y^2}$ orbitals, the
equivalent couplings along the $x$ and $y$ directions
favor other pairs of orbitals, i.e. $(d_{3x^2-r^2},d_{y^2-z^2}$ and $(d_{3y^2-r^2},d_{z^2-x^2})$.

\section{Compass model}
\label{compass}

A simple-looking spin model with similar properties (sometimes called the ``compass'' model) was introduced in 1982 in Ref.~[\onlinecite{KugelKhomskii}]:
\begin{equation}
H_{comp} = J \sum_{i} \left(S_{i}^xS_{i+x}^x + S_{i}^yS_{i+y}^y +
S_{i}^zS_{i+z}^z  \right)
\label{eq:compass}
\end{equation}
It is very similar to the model Eq.~(\ref{isocompass}), but while the three components of the pseudospin, $I^x$, $I^y$, and $I^z$  (see Eq.(\ref{Ioperators}), obey the condition $I^x + I^y + I^z = 0$ , the spin projections $S^\gamma$, $\gamma = x,y,z$ are independent operators.

Consider the model Eq.(\ref{eq:compass}) with $J < 0$ (ferromagnetic interactions), in which case the angular dependence of the spin exchange is similar to that of the conventional dipole-dipole interaction, and indeed, its properties can be qualitatively understood by considering  the lattice of ``compasses'' (or magnetic needles). The complications arising in this model are immediately clear from this analogy.
If, instead of the square lattice, we consider a one-dimensional raw of such compasses in the $z$ direction, it is clear that the magnetic needles would order "head to tail" parallel to the $z$ axis. However, in a row along the $x$ axis, spins would prefer to order in the $x$ direction. In a square and in a cubic lattice this leads to a strong  frustration, which in principle can destroy an order altogether. Such claim was indeed made by Glass and Lawson for a similar model with real dipole-dipole (long-range) interactions   \cite{GlassLawson}: using the Bogolyubov inequality they proved the absence of a long-range magnetic ordering at nonzero temperatures in two and three dimensions. Interestingly enough, the situation here is in a sense opposite to that in conventional systems: while in standard models with a continuous symmetry, like the Heisenberg model, the order is destroyed by fluctuations in one- and two-dimensional cases, but can exist in three-dimensional systems, here the order survives in one dimension, but is broken for $d = 2$ and $d=3$. This can be traced back to different symmetry properties of the two types of models: in contrast to the Heisenberg model, which has a continuous symmetry and, correspondingly, Goldstone modes, in Eqs. (\ref{isocompass}) and (\ref{eq:compass}) the interaction for each pair has an
Ising character. However, while in Heisenberg-like models
the spin and coordinate spaces are largely independent, in the model Eq.(\ref{eq:compass}) a rotation in the spin space is
inherently related to the corresponding rotation in the
coordinate space. This seems to ``restore'' the symmetry of the
system: if we assume a ferromagnetic  ordering for $d = 2,3$ and calculate the spin-wave excitation spectrum, we find that the
spectrum is gapless despite the discrete Ising type of interactions for each pair. Moreover, this spectrum turns out to be one-dimensional for a square lattice and two-dimensional for a cubic lattice:
\begin{equation}
\omega_{2d}^2 = J^2S^2 \left(1 - \cos k_x \right)
\end{equation}
and
\begin{equation}
\omega_{3d}^2 = J^2S^2\left(1- \cos k_x \right) \left(1
-\cos k_y\right)
\end{equation}
so that the corresponding fluctuations diverge at all
temperatures for $d = 2$, and at any $T>0$ for $d = 3$ case. This
agrees with the fact that on the mean-field level there are many
equivalent degenerate ground states in our system: apart from  ferromagnetic states with an arbitrary orientation of spins ${\bf S}$, there also states with ferromagnetic chains e.g. along the $z$ axis, with spins either parallel or antiparallel to the $z$-direction and no correlations between orientations of spins in different chains.

What is the true ground state of this model, is still an open problem. One possibility is that beyond the mean-filed approximation  the order-from-disorder scenario \cite{Villain,Shender} selects a particular ordered state \cite{KhaliullinOudovenko}. Another alternative is that the spin  fluctuations are so strong that they destroy a long-range order, leading to a (spin- or orbital-) liquid state \cite{Nussinov}. In any case, we see that properties of the models Eqs. (\ref{isocompass}) and (\ref{eq:compass}) are rather unusual.

\section{Exchange interactions in the degenerate Hubbard model}

We would like now to return to JT systems and discuss the purely electronic contribution to exchange interactions in these  materials. The calculation of the exchange Hamiltonian, resulting from the virtual hopping of electrons on neighboring sites, is usually done using the Hubbard model describing electrons on degenerate $d$-levels \cite{KugelKhomskii2,KugelKhomskii}
\begin{equation}
H = - \sum_{ij\alpha\beta\sigma} t_{ij}^{\alpha\beta} d^{\dagger}_{i\alpha\sigma}
d_{j\beta\sigma} + \sum_{i, \alpha \neq \beta} U_{\alpha\beta} n_{i\alpha\sigma}
n_{i\beta\sigma} - J_H \sum_i {\bf S}_{i1} \cdot {\bf S}_{i2}
\label{Hubbard}
\end{equation}
Here $i,j$ are site indices, $\sigma$ denotes spin projection, and $\alpha,\beta = 1,2$ are orbital indices. The second term in Eq.(\ref{Hubbard}) describes the on-site Coulomb repulsion  between electrons, which depends on the orbital occupation, and the last term is the Hund's rule exchange interaction between electrons  occupying two different orbitals on the same site. For an isolated ion, we have $U_{11} - U_{22} = 2J_H$. However, in crystals the direct Coulomb interactions $U_{\alpha\beta}$ are modified by the screening, while the Hund's rule exchange coupling $J_H$ remains unscreened. Therefore, $U_{\alpha\beta}$ and $J_H$ should be considered as independent parameters.

\begin{figure}
\centering
\includegraphics[width=5cm]{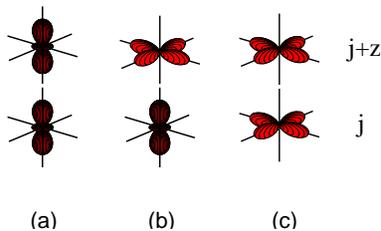}
\caption{\label{180} The overlap between $d$-orbitals of TM ions in the $180^\circ$-exchange.}
\end{figure}

For one electron or one hole per TM site, and in the strong coupling limit $U \gg J_H \gg W$, where $W$ is the electron band width, an effective Hamiltonian describing the coupled spin and orbital degrees of freedom can be obtained using the perturbation theory in $t/(U,J_{H})$ \cite{KugelKhomskii2}. The hopping matrix
elements $t_{ij}^{\alpha\beta}$ are spin-independent (spin is conserved in the process of the virtual electron hopping). As a result, the spin part of the exchange operator is rotationally invariant and has the standard Heisenberg form ${\bf S}_i \cdot {\bf S}_j$.  In contrast, the hopping amplitudes depend on the type of the initial and final orbital, as well as on
the spatial orientation of the pair of TM ions (see e.g.  Fig.\ref{180}), as a result of which the orbital part of the exchange Hamiltonian is strongly anisotropic.

For the pair of TM sites $j$ and $j+z$, the hopping only occurs between the two $d_{3z^2-r^2}$ orbitals (see Fig.\ref{180}a): the hopping between the $d_{3z^2-r^2}$ and  $d_{x^2-y^2}$  orbitals is zero by symmetry (see Fig.\ref{180}b), and the overlap between the two $d_{x^2-y^2}$  orbitals in the $z$ direction is very small (see Fig.\ref{180}c). Therefore, the usual type of exchange for this pair only occurs when the $d_{3z^2-r^2}$ orbitals ($T^z = \frac12$) are occupied on both sites. By the well-known Goodenough-Kanamori-Anderson  (GKA) rules, such an  exchange is antiferromagnetic, favoring the spin-singlet state of the two electrons (holes). This considerations allow us to immediately write an effective exchange Hamiltonian for this pair:
\begin{equation}
H_{j,j+z} = -\frac{4t^2}{U_{11}} \left(\frac12 + T^z_j\right)\left(\frac12 + T^z_{j+z}\right)
\left(\frac14 - {\bf S}_j \cdot {\bf S}_{j+z}\right),
\label{GKA}
\end{equation}
where $t$ is the hopping amplitude between the two  $d_{3z^2-r^2}$ orbitals. The exchange Hamiltonian is proportional to the product of the projection operators on the orbital states $|3z^2 - r^2\rangle$ on the two TM sites and the projection operator on the spin singlet state with ${\bf S}_j \cdot {\bf S}_{j+z} = -\frac34$. The ground state of such a pair is a spin singlet with an orbital configuration shown in Fig.\ref{180}a, in contrast to the (spin-triplet) state shown in Fig.\ref{jt},  stabilized by JT stabilized by the electron-lattice interaction.
Due to the electron-hole symmetry of the orbitally degenerate Hubbard Hamiltonian Eq.(\ref{Hubbard}),  the effective Hamiltonians for one hole and one electron per TM site  have the same form. There is also another mechanism of orbital exchange due to the hopping of electrons on empty orbitals of neighboring sites and the Hund's rule ferromagnetic spin coupling \cite{KugelKhomskii}, but it is usually weaker than the antiferromagnetic interaction Eq.(\ref{GKA}) and we shall not discuss it here.

As in the derivation of Eqs.~(\ref{Hx}) and (\ref{Hy}), the exchange interaction for pairs of TM ions along the $x$ and $y$ directions can be obtained from  Eq.(\ref{GKA}) by the corresponding rotation in the orbital space, i.e., by substituting the operators $T^z$ in that equation by, respectively, the $I^x$ and $I^y$ operators. The resulting exchange Hamiltonian for the cubic lattice is sometimes called the Kugel-Khomskii (KK) model.  The coupled orbital and spin exchange, described by Eq.(\ref{GKA}) and the corresponding expressions for $H_{j,j+x}$ and $H_{j,j+y}$, has the same spatial anisotropy as the lattice-mediated orbital interactions  Eq.(\ref{isocompass}), and we face again the problem of frustrated orbital ordering. It was even suggested that the coupling between orbitals and spins in Eq.(\ref{GKA}) makes it only worse: it gives anomalously soft mixed spin-orbital excitations, resulting in a spin-orbital liquid ground state \cite{FeinerOlesZaanen,OlesFeinerZaanen}. Other authors, however, argued that this coupling actually helps to lift the frustration, resulting in orbitally and magnetically ordered states, in which spins form antiferromagnetic chains (as in the cubic perovskite KCuF$_3$), stabilized by large quantum spin fluctuations \cite{KhaliullinOudovenko}.  It seems that the only way to clarify this controversy is to perform a direct numerical simulation of the model Eq.(\ref{GKA}), but the latter is complicated by the sign problem (which again is a consequence of frustration).

\section{$90^{\circ}$-exchange}
\label{90}

So far, we only discussed a simple cubic lattice of TM ions and showed that even in that case orbital interactions are frustrated. There are also many Jahn-Teller materials with  lattices that cannot be naturally subdivided into two
sublattices of TM ions, e.g. a two-dimensional triangular and Kagome lattice, or a pyrochlore lattice of corner-shared tetrahedra, e.g. the B-sublattice of spinels. An antiferromagnetic spin ordering in materials with such lattices is known to be strongly frustrated \cite{Ramirez} and it is interesting to see,  if such `geometric' effect can enhance the frustration of orbital interactions.

\begin{figure}
\centering
\includegraphics[width=5cm]{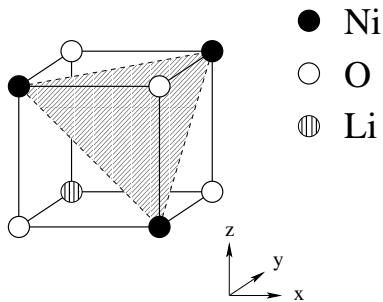}
\caption{\label{nili} The crystal structure of LiNiO$_2$. The sublattice of Ni ions consists of triangular [111] layers (shaded).}
\end{figure}

The typical example of such a material is LiNiO$_2$, in which the low-spin Ni$^{3+}(t_{2g}^6e_g^1)$ ions with spin
$S=\frac12$ and doubly degenerate orbitals ($T = \frac12$),
occupy two-dimensional triangular [111] layers (see Fig.\ref{nili}). While in perovskites the angle between metal-oxygen bonds connecting neighboring TM ions is $180^\circ$, in LiNiO$_2$ this angle is $90^\circ$. This difference has important consequences for the spin and orbital exchange \cite{MostovoyKhomskii}.

The actual exchange in TM oxides occurs not due to a
direct $d-d$-hopping, as described by Eq.(\ref{Hubbard}), but involves the hopping from oxygens to TM sites and intermediate states with holes on oxygen $2p$-orbitals. Thus we have to start from the d-p model:
\begin{eqnarray}
H_{dp} &=& \sum_{im\sigma}\varepsilon_p p_{im\sigma}^\dagger p_{im\sigma} +  \sum_{jm\sigma}\varepsilon_d d_{j\alpha\sigma}^\dagger d_{j\alpha\sigma} - \sum_{m\alpha ij\sigma} t_{ij}^{m\alpha} \left(p_{im\sigma}^\dagger d_{j\alpha\sigma} + d_{j\alpha\sigma}^\dagger p_{im\sigma} \right) \nonumber \\
 &+& \sum_{i m m'} U_{mm'} n_{im\sigma}
n_{im'\sigma} + \sum_{j \alpha \beta} U_{\alpha\beta} n_{j\alpha\sigma}
n_{j\beta\sigma} \nonumber \\
&-&  J_{H}^p \sum_{i,m \neq m'} {\bf S}_{im} \cdot {\bf S}_{im'} - J_{H}^d \sum_{j,\alpha \neq \beta} {\bf S}_{j\alpha} \cdot {\bf S}_{j\beta}
\label{Hdp}
\end{eqnarray}
Here $m=x,y,z$ and $\alpha$ are the orbital indices of, respectively, oxygen $2p$-orbitals and TM $d$-orbitals, and the Hamiltonian of the model includes the direct (Hubbard) and the exchange (Hund's rule)  Coulomb interactions on both the oxygen and TM ions.

In the Mott-Hubbard (MH) region of the general Zaanen-Sawatzky-Allen diagram \cite{ZaanenSawatzkyAllen} (see also \cite{KhomskiiSawatzky}), i.e. when the charge transfer gap
$\Delta \gg U \gg W$ ($\Delta$  is $\varepsilon_p - \varepsilon_d$ for one hole and $\varepsilon_d + U_{dd} -
\varepsilon_p$ for one electron per TM site, and $W$ is the band width), the model Eq.(\ref{Hdp}) can be reduced to the effective Hubbard model for $d$-electrons Eq.(\ref{Hubbard}) with $t_{dd} =   \frac{t_{pd}^2}{\Delta}$. On the other hand, in the charge transfer (CT) insulator regime,  $U_d \gg \Delta \gg W$, one finds new orbital exchange terms that turn out to be more important than those obtained within the Hubbard model Eq.(\ref{Hubbard})  (see the discussion below).

\begin{figure}
\centering
\includegraphics[width=7cm]{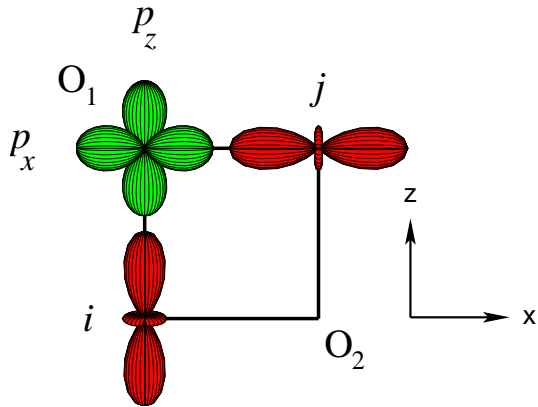}
\caption{\label{plaquette} The overlap between the
$d$-orbitals of the TM sites $i$ and $j$ and with two different $p$-orbitals of oxygen $O_1$ for the $90^\circ$-exchange.}
\end{figure}

In any case, exchange interactions for the $90^\circ$-angle between the metal-oxygen bonds (see Fig.\ref{nili}) are very different from Eq.(\ref{GKA}) for the $180^\circ$-angle. In the $90^\circ$-geometry the orbitals of one TM (e.g. site $i$ in Fig.\ref{plaquette}) overlap with the $p_z$-orbital of the oxygen O1 (see Fig.\ref{plaquette}), whereas the $d$-orbitals of the ion $j$ overlap with the orthogonal $p_x$-orbital of the same oxygen. Due to the  orthogonality of these two oxygen orbitals, it is impossible to transfer a $d$-electron from one TM ion to another via oxygens, and this is the principal difference from the $180^\circ$-exchange. The only processes allowed are the transfers of oxygen electrons to the site $i$ (from the $p_z$-orbital) followed by the transfer of another electron from the $p_x$-orbital of this oxygen to the the TM site $j$. If we first ignore the Hund's rule coupling on oxygens, these processes are independent of spins of $d$-electrons, i.e., they do not result in a spin exchange. However, they do depend on the orbital occupation of the two TM ions, resulting in an orbital exchange. Therefore, for this geometry, the spin and orbital interactions are decoupled, the orbital
exchange being much stronger than the spin one (the latter $\propto  J_H / \Delta \ll 1$). On the other hand, in the
KK model Eq.(\ref{GKA}) the strongly coupled spin and orbital exchanges are of the same order. Thus, the KK model, obtained for the specific case of the MH insulators with the $180^\circ$ metal-oxygen-metal bonds, should not be used uncritically for other situations.

These general considerations are confirmed by detailed
calculations \cite{MostovoyKhomskii}. For instance, for the plaquette formed by two oxygens and two TM ions on the sites $i$ and $j$ in the $zx$ plane (see Fig.\ref{plaquette})  the orbital exchange interaction is given by
\begin{equation}
H_{T} = J_T \,\left[\left(\frac{3}{2} -
I_i^x\right)\! \left(\frac{3}{2} - I_j^z\right)\!
+\! \left(\frac{3}{2} - I_i^z\right)\!
\left(\frac{3}{2} - I_j^x\right)\right],
\label{HT}
\end{equation}
where
\begin{equation}
J_T =-\frac{4t_{pd}^4}{\Delta^2(2\Delta+U_p)} + \frac{2
t_{pd}^4}{\Delta^3} = \frac{2 t_{pd}^4 U_p}{\Delta^3 (2\Delta +
U_p)}, \label{eq:J_T}
\end{equation}
and the coupled spin-dependent exchange is
\begin{equation}
H_{TS} = - \frac{2 J^p_H}{(2\Delta +
U_p)} H_T
\left(\frac{3}{4} + ({\bf S}_1 {\bf S}_2)\right).
\label{eq:HTS}
\end{equation}

Several points have to be stressed here. First of all, we indeed
see that the orbital exchange is stronger than the spin
one (usually $J^p_H \sim 0.6-0.8$eV and both $U_p$ and $\Delta$ are $\sim 4-6$eV, so that $J_H^p < U_p,\Delta$). The spin exchange only appears due to the Hund's rule coupling (in this case on oxygen) and is always {\it ferromagnetic}, independent of the
orbital occupation. Therefore, the spin ordering on this triangular lattice is not frustrated. This, in particular, invalidates the attempts to describe LiNiO$_2$ as a spin liquid of the RVB type \cite{Hirakawa}.

The second point is the presence of two terms with opposite signs
in the expression for orbital exchange Eq.(\ref{HT}). The origin of these terms is rather interesting. The negative term resembles those obtained in the Hubbard model: it
describes the virtual process with two holes on one oxygen in the intermediate state, which is why it contains $2\Delta+U_p$ in the denominator. The positive term, however, is of a different nature  - it remains nonzero even when $U_p \rightarrow \infty$. Its origin can be understood as follows: the virtual hopping of $d$-electrons from a TM site to $6$ surrounding oxygens results in an energy decrease. However, each oxygen belongs to two different octahedra. For infinite $U_p$ this leads to blocking of some of these charge fluctuations: when e.g. the oxygen $O_1$ in Fig.\ref{plaquette} is occupied by the hole from the site $i$, it blocks the virtual hole transfer from the site $j$. This blocking reduces the total energy gain due to the charge fluctuations and gives rise to an effective orbital exchange interaction, since an amount by which the energy is reduced  depends on orbital occupations on the sites $i$ and $j$. This mechanism of exchange is similar to the Casimir interaction between two metallic plates in vacuum.

While the Hubbard model gives an effective interaction, which respects the electron-hole symmetry, in the d-p model this symmetry is lost. In particular, for one hole per TM site the effective orbital interaction the form
\begin{equation}
H_{T}^{(hole)} = J_T \,\left[\left(\frac{3}{2} -
I_i^x\right)\! \left(\frac{3}{2} - I_j^z\right)\!
+\! \left(\frac{3}{2} - I_i^z\right)\!
\left(\frac{3}{2} - I_j^x\right)\right],
\label{HThole}
\end{equation}
(cf. Eq.(\ref{HT}) for one electron per site). In this case the effect of blocking is minimized when  the $d$-hole orbital on at least one site is not directed towards the common oxygen, e.g. $d_{3z^2-r^2}$-orbital on the site $j$ in Fig.\ref{180}b   and $d_{x^2-y^2}$-orbital on the site $j+z$. Therefore, the orbital interaction Eq.(\ref{HThole}) favors hole orbitals directed away from the common oxygens, in contrast to the situation in MH insulators (see Eq.(\ref{GKA})), for which the energetically most favorable occupation is shown in Fig.\ref{180}a.

\begin{figure}
\includegraphics[width=7cm]{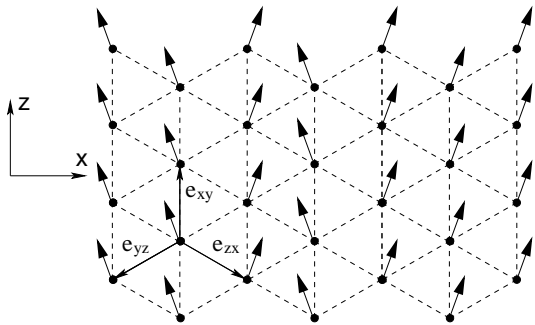}
\caption{\label{triang}  A disordered
mean-field ground state, in which the pseudospins form lines
parallel to the unit vector ${\bf e}_{xy}$, such that
$\langle T^z_j \rangle$ is the same on all lattice sites,
while the sign of $\langle T^x_j \rangle$ varies
arbitrarily from line to line.}
\end{figure}

If we now consider the exchange Hamiltonian for the whole triangular lattice, we find out that here too the interactions in different directions compete. The outcome is again that in the
mean-field approximation the ground state is strongly degenerate. The orbital exchange on a triangular lattice  is strongly
frustrated and has a large number of mean-field ground
states, including ferro-orbital states, in which $\langle {\bf T}_j \rangle
= T {\bf m}$ on all lattice sites, where ${\bf m}$ is an
arbitrary unit vector in the $(T^x,T^z)$-plane.
Furthermore, there are disordered ground states, which can
be obtained from the ferro-orbital states by, e.g., inverting the
sign of $\langle T^x_j \rangle$ on an arbitrarily selected
set of lines parallel to the lines with the $I^x I^y$ type of exchange (see Fig.~\ref{triang}). Such states are ordered along the
lines, but there are no long-range correlations
between the $x$-projections of pseudospins in the transverse
direction (as in the compass model discussed in Sec.~\ref{compass}). By circular permutations of the $x$, $y$, and
$z$ indices one can obtain similar states, which are only
ordered along the two other sets of lines in the triangular lattice.

The spectrum of orbital excitations for, e.g., $\langle {\bf T} \rangle = T {\hat z}$, calculated in the (pseudo)spin-wave approximation for $T \gg 1$, is given by
\begin{equation}
\omega_{\bf q} = 3 \sqrt{2} T J_T  \left| \sin\frac{q_z}{2} \right|,
\label{eq:1d}
\end{equation}
As in the case of the compass model discussed in Sec.~\ref{compass}, the excitation spectrum is gapless and it is independent of one of the projection of momentum ($q_x$). This one-dimensional spectrum leads to divergent
fluctuations, to remove which, one has to go beyond the (pseudo)spin-wave approximation and take into account interactions between the orbital excitations. One obtains then that the order-from-disorder mechanism chooses 6 particular ferroorbital ground states, e.g. $|3z^2 - r^2\rangle$ (all $\langle T^z_i \rangle = +\frac12$) \cite{MostovoyKhomskii} and that the spectrum of elementary excitations acquires the gap $\propto J_T \sqrt{T}$, makes the orbital fluctuations finite.

We note that the `geometric' frustration of the triangular lattice plays an important  role in this story. One can show \cite{MostovoyKhomskii2} that, in contrast to the ``compass'' model, an ordering of $e_g$ orbitals on a square lattice is not frustrated. Thus in two dimensions, both the anisotropy of the orbital exchange and the special geometry of the triangular lattice are necessary for frustration. In a similar way one can treat other types of `frustrated' lattices. For instance, effective interactions  Eqs.(\ref{HT}) and (\ref{eq:HTS}) can be used to describe exchange on the pyrochlore lattice of TM ions,  since the local geometry of metal-oxygen-metal bonds is the same as in LiNiO$_2$. Also the  orbital excitations for the $d_{3z^2 - r^2}$ ferro-orbital state on the pyrochlore lattice only propagate along two disconnected sets of bonds with the
$I^x I^y$-type of exchange, similarly to the case of the
triangular lattice.

Applying the d-p model to systems with the $180^\circ$-exchange path, one can show that the `orbital Casimir' interactions also exist in these systems \cite{MostovoyKhomskii2}. They have the form similar to the phonon-mediated interactions Eq.(\ref{isocompass}) and modify the standard KK description.
In charge transfer insulators these terms dominate the exchange, resulting in the decoupling of orbital and spin orderings  and forcing orbitals to order at higher temperatures than spins, provided that the frustration of the orbital interactions is somehow lifted.

\section{Summary}

Summarizing, we have discussed orbital interactions and frustration of orbital ordering in different physical situations. We showed that due to the directional nature of the hopping amplitudes the orbital exchange is frustrated even for simple (perovskite) lattices. Frustration is therefore a generic property of orbital degrees of freedom. If in addition
we are dealing with the systems with the `geometric' frustration,
the ground state degeneracy becomes even stronger. The resulting type of the orbital, and spin, structure of the ground state can then be determined either by small corrections to the exchange interactions lifting this degeneracy, e.g. the Hund's rule coupling or small ``external'' splitting of degenerate $e_g$-levels due to deviations from the perfect cubic symmetry. Another possibility is an ordering due to the order-from-disorder mechanism, discussed above. We note, however,  that this semiclassical mechanism, strictly speaking, works for large values of pseudospin $T \gg 1$, when quantum fluctuations are relatively small. Since in reality $T=\frac12$, strong quantum fluctuations may still result in a disordered ground state. Such a quantum orbital liquid remains an intriguing possibility.

The situation with other types of orbitally-degenerate systems,
e.g. materials with three-fold degenerate $t_{2g}$-orbitals, may have its own peculiarities \cite{KugelKhomskii3}. First of all, due to a larger degeneracy and a weaker coupling to the lattice, one may expect stronger frustration  and stronger quantum effects, which could result in an orbital-liquid ground state \cite{KhaliullinMaekawa}. On the other hand, as we already mentioned in the Introduction, in  $t_{2g}$-systems the spin-orbital interaction may play an important role, which  may remove the orbital
degeneracy \cite{KugelKhomskii3}. Furthermore, in frustrated lattices (of the type
considered in Sec.~\ref{90}) of  $t_{2g}$-ions, one has to take into account not only the exchange between the ions via oxygens, but also direct the  $t_{2g}-t_{2g}$ overlaps across the TM-oxygen plaquettes (see Fig.\ref{plaquette}). What would be the resulting orbital structure with all these effects taken into account, is still an open problem. We see, that the orbital physics, especially in materials with complex lattices, is very rich and can still produce many surprises.

This work is supported by the MSC$^{plus}$ program. The financial support of the British Council and the Trinity College, Cambridge is gratefully acknowledged.

\end{document}